\documentclass[twocolumn,showpacs,pre,aps,amssymb,amsmath]{revtex4}
\usepackage{graphicx}
\newcommand{\be}{\begin{equation}}
\newcommand{\ee}{\end{equation}}
\newcommand{\bd}{\begin{displaystyle}}
\newcommand{\ed}{\end{displaystyle}}
\newcounter{abcd}

\begin{document}
\title{Chaos and flights in the atom-photon interaction in cavity QED}

\author{S.V. Prants}
\affiliation{Laboratory of Nonlinear Dynamical Systems,
V.I. Il'ichev Pacific Oceanological Institute
of the Russian Academy of
Sciences,
2690041 Vladivostok, Russia}
\author{M. Edelman}
\affiliation{Courant Institute of Mathematical Sciences,
New York University,
251 Mercer St., New York, NY 10012, USA}
\author{G.M. Zaslavsky}
\affiliation{Courant Institute of Mathematical Sciences,
New York University,
251 Mercer St., New York, NY 10012, USA\\
Department of Physics, New York University,
2-4 Washington Place, New York, NY 10003, USA}

\date{\today}
%\maketitle
\begin{abstract}
We study dynamics of the atom-photon interaction in 
cavity quantum electrodynamics (QED), considering
a cold two-level atom in a single-mode high-finesse
standing-wave cavity as a nonlinear Hamiltonian system with three
 coupled degrees of freedom: translational, internal atomic,
and the field. The system proves to have
different types of motion including L\'{e}vy flights and chaotic
walkings of an atom in a cavity. It is shown that the
translational motion, related to the atom recoils,
is governed by an equation of
 a parametric nonlinear  pendulum with a frequency  modulated
by the Rabi oscillations. This type of dynamics
is  chaotic
with some width of the stochastic layer that is estimated analytically.
The width is fairly
small for realistic values of the control parameters, the
normalized detuning $\delta$ and atomic recoil frequency $\alpha$.
It is demonstrated how the atom-photon
dynamics  with a  given value of  $\alpha$ depends on the values of
$\delta$ and initial conditions. Two types of L\'{e}vy flights, one
corresponding to the ballistic motion of the atom and another
 one corresponding to small oscillations in a
potential well, are found. These flights influence
statistical properties of the atom-photon interaction such  as
distribution of Poincar\'{e} recurrences and moments of the atom
position $x$. The simulation shows different regimes of motion,
from slightly abnormal diffusion with $<x^2> \sim \tau^{1.13}$  at
$\delta =1.2$ to a superdiffusion with $<x^2> \sim \tau^{2.2}$  at
$\delta =1.92$ that corresponds to a superballistic motion of
 the atom with an acceleration. The obtained results can be
used to find new ways to manipulate atoms, to cool and trap
them by adjusting the detuning $\delta$.
\end{abstract}
\maketitle

\section{Introduction}

Cavity quantum electrodynamics is a rapidly developing field of
physics studying the interaction of atoms with photons in high-finesse
cavities in a wide range of electromagnetic spectrum, from microwaves to
visible light, in such conditions under which both atoms and fields may
manifest their quantum nature (for reviews on quantum and atom optics, see
\cite{B,ASM}). Modern experiments in cavity QED have achieved the
exceptional circumstance of strong atom-field coupling for which the strength
of the coupling exceeds both the atomic dipole decay and the cavity field
decay providing manipulations with single atoms and photons
\cite{Y99,HLD,EMK,MF99,PFM}.
Trapped atoms and ions, interacting with laser fields in
the regime of the strong coupling,
have been used not only to study fundamentals of quantum mechanics
\cite{W98,MM96}
but also for applications
in the rapidly growing fields of quantum computing, quantum communication
(see, for example,
\cite{CZ95,MM95}), quantum chaos
\cite{Z81,H91}, and decoherence
\cite{Z91,GJ96}.

A special comment should be related to the notion of chaos that we
use in the paper. The real system is quantum and the quantum chaos
per se does not  exist.
The notion ``quantum chaos'' just refers to the behavior of quantum
systems whose classical counterparts behave chaotically.
Through the paper the notion ``chaos''
will be used only in that meaning since all our consideration will 
be semiclassical. The key issue of quantum chaos is the correspondence
between the classical and quantum pictures of chaotic dynamics.
Chaotic dynamics  in the atom-photon physics appeared in
the paper \cite{Haken}, where the semiclassical model of a single-mode,
resonant, and homogeneously broadened laser, considered as an open
dissipative system, has been shown to be equivalent to a Lorentz-type strange
attractor, and the paper
\cite{BZT}, where Hamiltonian semiclassical chaos has been shown to arise
in the Dicke model with non-resonant terms describing the interaction of
an ensemble of identical two-level atoms with their own radiation field in
an ideal resonant cavity. The fully quantum version of the latter model
has been considered in \cite{BBZ}. It was shown in [18]
that in a parameter range, corresponding to classical chaos,
the evolution of the system becomes essentially quantum after the so-called
``breaking time''  $\tau_{\hbar}$ \cite{BZ78}
which obeys in this regime the logarithmic  law
$\tau_{\hbar} \sim \lambda^{-1}  \ln({\rm const}/\hbar)$, 
where $\lambda$ is the maximal Lyapunov exponent. 
All results presented here as chaotic dynamics are
valid up to $t \lesssim \tau_{\hbar}$.

The physical mechanism of semiclassical chaos in the Dicke model is tied
to virtual transitions which are described by non-resonant (or
counter-rotating) terms in the Dicke Hamiltonian. They are small under usual
conditions. Trying to find another mechanism of local instability in the
Hamiltonian atom-photon dynamics without pump and losses, the authors
of \cite{PK97,PKK99}
proposed the semiclassical model with atoms moving through a standing-wave
cavity in a direction along which the cavity sustains a
space-periodic field. The standing wave modulates the atom-field coupling
providing in a certain range of the system's parameters intermittent Rabi
oscillations. Another way to change the atom-field coupling is
modulating the cavity length, studied in \cite{IKP}.
New effects in the model with moving atoms may arise beyond the simple
semiclassical approximation. Chaotic vacuum Rabi oscillations, a new kind of
reversible spontaneous emission, have been shown
\cite{P00,PK00} to occur in the model with interatomic quantum correlations.

Experiments to study the quantum dynamics of classically chaotic systems
for the atom-photon interaction in cavities and traps have been
intensively studied with cold atoms in a phase-modulated standing wave
\cite{MR94} and in an amplitude-modulated standing wave \cite{HT}
(following to the proposition of \cite{GSZ}), and in a pulsed standing wave
\cite{MR95,AG}.
Cold sodium or cesium atoms, which are kicked by a periodically pulsed
standing wave of far-detuned light, is an excellent experimental realization
of a paradigm model of quantum chaos, a $\delta$-kicked quantum rotor.
Dynamical localization, that was observed in the atomic momentum
distribution, is the quantum suppression of the classical momentum diffusion.
A typical underlying phase-space structure of classical chaotic systems
consists of stochastic webs, islands, and chains of islands embedded in the
stochastic sea. The chaotic motion occupies a certain area in
phase space. Because of islands and their boundaries, typical behavior of chaotic
systems can be intermittent with long (quasi)regular oscillations (the so-called
L\'{e}vy flights) interrupted by the chaotic pieces of trajectories. This
intermittency leads to the anomalous diffusion with
L\'{e}vy distribution functions or a similar one,
which have power-wise tails (for reviews on
L\'{e}vy processes in physics   see \cite{SZK}, where the term
``strange kinetics'' was coined, and \cite{Z98}). It was found with kicked
cold cesium atoms \cite{K0}
that for certain pulses amplitudes, where the respective classical analog
may exhibit anomalous diffusion, the momentum distributions were not
exponentially localized
for the time of observation (see also
\cite{BS99,AI99,RA01}).
It should be mentioned that anomalous diffusion and
L\'{e}vy flights
have been even earlier found with cold atoms and employed in a subrecoil
laser cooling scheme
\cite{BB94,RB95,SL99}.

In experiments with $\delta$-kicked atoms, the detuning between the optical
and atomic transition frequencies is large (relative to the natural
line-width), so the probability is small to find an atom, initially prepared 
in the ground state, in the excited state,
and the excited state amplitude can be adiabatically eliminated
\cite{GSZ}.
In this approximation, an effective Hamiltonian is that of a driven
nonlinear oscillator with 3/2 degrees of freedom. Generally
speaking, the atom-photon interaction in a high-finesse
cavity is, mainly,  the
interaction between internal (electronic) and external (motional) atomic
degrees of freedom and the cavity field. A corresponding  one-dimensional model,
including the interaction of all those degrees of freedom, has been
introduced in papers \cite{PK01,PS01} in the context of Hamiltonian chaos.
In this paper, we use a slightly
generalized version of that model with three degrees of freedom to study
the effects of chaos and L\'{e}vy flights in the strongly-coupled atom-field
system.

In \setcounter{abcd}{2}Sec.~\Roman{abcd} the semiclassical
basic equations in the Heisenberg representation are derived
in the form of six coupled nonlinear equations with 
two control parameters, the normalized
detuning $\delta$ and atomic recoil frequency $\alpha$. Taking into account
that the frequency of the atomic (and field) Rabi oscillations
is much more larger than the frequency of translational
oscillations, reduced Bloch-like equations of the atom-field
internal motion are derived and solved in \setcounter{abcd}{3}
Sec.~\Roman{abcd}. The motion of the center of the atom mass
is governed by a single equation of a parametrically perturbed
pendulum. This dynamics generates
a stochastic layer with an exponentially small width.
In  \setcounter{abcd}{4}Sec.~\Roman{abcd}
we simulate
the basic set of  equations at the fixed value
of the normalized recoil frequency $\alpha =10^{-3}$ corresponding to
a light atom in a microcavity with realistic parameters
\cite{Y99,HLD,EMK,MF99,PFM}.
The maximal Lyapunov exponents and Poincar\'{e}
sections are calculated and it is found that the
detuning $\delta$ is a crucial parameter in transition to chaos.
Statistical properties of the atom-photon interaction are
considered in \setcounter{abcd}{5}Sec.~\Roman{abcd}. We give an
evidence of two types
of L\'{e}vy flights of an atom, one corresponding to
an almost linear dependence of the atomic position $x$ on time
(superdiffusive and superballistic regime),
and another one corresponding to small regular oscillations
of the atom in the potential well. The L\'{e}vy flights influence strongly
 such statistical properties of atoms and photons as distribution
of Poincar\'{e} recurrences and moments of $x$. The distribution
of recurrences and time-evolution of the moments depend
on the value of the detuning $\delta$ demonstrating different regimes
from an almost normal diffusion to a superdiffusion. In \setcounter{abcd}{6}
Sec.~\Roman{abcd} we discuss briefly
ways to manipulate the atomic motion by varying control
parameters and initial internal atomic states. It is possible,
in particular, to cool and trap atoms by adjusting the detuning.

\section{Basic equations}

The basic model of interaction of radiation with matter describes the
energy exchange between a two-level atom and a single mode of the
quantized radiation field in an ideal lossless cavity
\cite{JC}.
In general, this interaction should involve not only the internal atomic
transitions and field states but also the center-of-mass motion of the atom.
With the recoil effect to be included into consideration, the standard
Jaynes-Cummings Hamiltonian can be extended as follows:
\be
H = \frac{\hat{p}^2}{2m} + \hbar\omega_a \hat{s}_z +
\hbar\omega_f \hat{a}^+ \hat{a} - \hbar\Omega_0
(\hat{a}^+ \hat{s}_- + \hat{a}\hat{s}_+ ) \cdot\cos k_f \hat{x}
\ , \label{2.1}
\ee
where $\hat{x}$ and $\hat{p}$ are the atomic position and momentum operators,
respectively. Transitions between two electronic states, separated by
the energy $\hbar\omega_a$, are described by the spin operators 
with the commutation relations,
$[\hat{s}_+ ,\hat{s}_- ] = 2\hat{s}_z$ and $[\hat{s}_z, \hat{s}_{\pm}]= 
{\pm}\hat{s}_{\pm}$. The photon annihilation and creation operators with
the commutation rule $[\hat{a},\hat{a}^+ ] = 1$ characterize a selected
mode of the radiation field of the frequency $\omega_f$ and the wave number
$k_f$ in a lossless cavity of the Fabry-Perot type. The parameter $\Omega_0$
is the amplitude value of the atom-field dipole coupling which depends on
the position of an atom inside a cavity.  As it is usually adopted in cavity 
QED, we write down the Heisenberg equations for the
external atomic operators, $\hat{x}$ and $\hat{p}$, and for slowly-varying 
amplitudes of the field and spin operators $\hat{a}(t)=\hat{a} 
\exp(-i\omega_ft),\quad \hat{a}^+(t)=\hat{a}^+ \exp(i\omega_ft),\quad 
\hat{s}_{\pm}(t)=\hat{s}_{\pm}\exp(\pm i\omega_ft)$ and $\hat{s}_z(t)=
\hat{s}_z$
\begin{equation}
\begin{array}{lll}
\frac{d}{dt} \hat{x} = \frac{\hat{p}}{m}\ ,\\
\frac{d}{dt} \hat{p} = -\hbar k_f \Omega_0
(\hat{a}^+ \hat{s}_- + \hat{a}\hat{s}_+ ) \sin k_f \hat{x} \ ,\\
\frac{d}{dt} \hat{s}_+ = -i(\omega_f -\omega_a )\hat{s}_+ + 2i\Omega_0\hat{a}^+
\hat{s}_z \cos k_f \hat{x} \ ,\\
\frac{d}{dt} \hat{s}_- = i(\omega_f -\omega_a )\hat{s}_- - 2i\Omega_0\hat{a} \
\hat{s}_z \cos k_f \hat{x} \ , \\
\frac{d}{dt} \hat{s}_z = -i\Omega_0
(\hat{a}^+ \hat{s}_- - \hat{a}\hat{s}_+ ) \cos k_f \hat{x} \ ,\\
\frac{d}{dt} \hat{a}^+ = -i\Omega_0 \hat{s}_+ \cos k_f \hat{x}\ ,\\
\frac{d}{dt} \hat{a} = i\Omega_0 \hat{s}_- \cos k_f \hat{x}\ .
\end{array}
\label{2.2}
\end{equation}
To avoid cumbersome notations in Eqs. (\ref{2.2}) we use for the amplitudes the same 
notations as for the respective whole operators. 

In order to derive a tractable closed set of equations for expectation
values from the Heisenberg operator equations
(\ref{2.2}),
we use the semiclassical approximation. It means that all the operators and
their products in Eqs. (\ref{2.2})
are averaged over an initial quantum state, which is supposed to be a
product state of the translational, electronic, and field states. The
expectation values of all the operator products are factorized to the products
of the respective expectation values, e.g.,
$\langle (\hat{a}^+ \hat{s}_- + \hat{a}\hat{s}_+ ) \cdot\sin k_f \hat{x}
\rangle$
$= (\langle \hat{a}^+ \rangle \langle \hat{s}_- \rangle + \langle
  \hat{a} \rangle \langle\hat{s}_+ \rangle )
\sin (k_f \langle \hat{x} \rangle ) $.
By choosing the following dimensionless expectation values:
$x=k_f \langle \hat{x} \rangle$,
$p = \langle\hat{p} \rangle /\hbar k_f$,
$a_x = \frac{1}{2} \langle \hat{a} + \hat{a}^+ \rangle$,
$a_y = \frac{1}{2i} \langle \hat{a} - \hat{a}^+ \rangle$,
$s_x = \frac{1}{2} \langle \hat{s}_- + \hat{s}_+ \rangle$,
$s_y = \frac{1}{2i} \langle \hat{s}_- - \hat{s}_+ \rangle$,
as dynamical variables,
we finally get from Eqs. (\ref{2.2}) a nonlinear dynamical system
\begin{equation}
\begin{array}{lll}
\dot{x} = \alpha p \ ,\\
\dot{p} = -2 (a_x s_x + a_y s_y ) \sin x \ ,\\
\dot{s}_x = - \delta s_y + 2a_y s_z \cos x \ ,\\
\dot{s}_y =  \delta s_x - 2a_x s_z \cos x \ ,\\
\dot{a}_x = - s_y \cos x \ ,\\
\dot{a}_y =  s_x \cos x \ ,
\end{array}
\label{2.3}
\end{equation}
that describes the interaction between three degrees of freedom 
in the strongly coupled atom-field system, translational $(p,x)$,
electronic $(s_x ,s_y )$, and the field $(a_x ,a_y )$ ones. 
The dot in Eqs. (\ref{2.3}) denotes the derivative with respect to the
normalized time $\tau = \Omega_0 t$. The control parameters are
the normalized recoil frequency $\alpha = \hbar k_f^2 /m\Omega_0$ and the
normalized detuning between the frequencies of the field mode and the
electronic transition, $\delta  = (\omega_f - \omega_a )/\Omega_0$.
The system (\ref{2.3}) conserves the energy
\begin{equation}
W = \frac{\alpha}{2} p^2 -2 (a_x s_x + a_y s_y ) \cos x -
\delta s_z \ ,
\label{2.4}
\end{equation}
and it possesses two additional first integrals
\begin{equation}
s_x^2 + s_y^2 + s_z^2 = S^2 , \quad a_x^2 +a_y^2 + s_z = N \ .
\label{2.5}
\end{equation}
The first one is simply the conservation of the atomic probability,
and the second one is a conserved total number of excitations, which is
known to be a constant in the rotating-wave approximation. The equation 
of motion  for the atomic inversion $s_z$ is easily derived with the help of 
the integral $N$  
\begin{equation}
\dot{s}_z = 2 (a_x s_y - a_y s_x ) \cos x \ .
\label{2.6}
\end{equation}
The semiclassical approximation, we used for
(\ref{2.3}), means that  the atom as a
classical particle with external and internal states moves in a
self-consistent classical radiation field.

From the dynamical systems point of view equations (\ref{2.3})
represent a system with 3 degrees of freedom 
(one degree of freedom per a canonically
conjugate pair of the generalized momentum and coordinate)
in 6-dimensional phase space. Indeed, we reinsert Eq.(\ref{2.6})
for $\dot{s}_z$ into (\ref{2.3}). After that the system (\ref{2.3})
describes 3 degrees of freedom: the atom external coordinates
($x,p$), the atom internal coordinates ($s_x,s_y$), and the field 
coordinates ($a_x,a_y$).
There are 2 constrains: energy 
integral $E=H(p,x; s_x, s_y; a_x, a_y)$ and the spin integral $S$.  
The number of excitations $N$ should be used to determine $s_z$ as a function of
other variables. That means that full dynamics is defined in 4-dimensional
hyperspace and should have domains of chaos due to its nonintegrability.
One can say that the location of the domains of chaotic motion,
islands of regular dynamics, set of stationary points, and boundaries
define a topology of the system's flow. The topology is two-parametric 
$(E, S)$ and very complicated. Its description needs a separate investigation. 

\section{Reduced dynamics and the estimation of the stochastic layer width}

We can simplify further the basic equations
(\ref{2.3}) introducing the combined atom-field variables
\be
u=2(a_x s_x + a_y s_y ), \quad v = 2(a_y s_x - a_x s_y )
\label{3.1}
\ee
and using the integrals
(\ref{2.5}).
As a result, one arrives at the closed five-dimensional dynamical system
\begin{equation}
\begin{array}{lll}
\dot{x} = \alpha p \ ,\\
\dot{p} =- u \sin x \ ,\\
\dot{u} = \delta v \ ,\\
\dot{v} = - \delta u + 2(2N s_z - 3s_z^2 + S^2 ) \cos x \ ,\\
\dot{s}_z = - v \cos x \ ,
\end{array}
\label{3.2}
\end{equation}
which generalizes the corresponding equations of the paper
\cite{PS01}
(see Eqs. (3) therein which were derived in the limit of large $N$).
It is obvious from Eqs. (\ref{3.2})
that at exact resonance, $\delta = 0$, the slow translational variables
$x$ and $p$ are separated from the fast atom-field   $u,v$, and $s_z$,
and the system (\ref{3.2})
becomes integrable. At $\delta =0$,  the atom moves in a spatially
periodic optical potential $U = -u(0) \cos x$ with $u(0) = u(\tau =0)$ =
const, and its center-of-mass motion satisfies the
pendulum equation $\ddot{x} + \alpha u(0) \sin x = 0$. It is easy to find
that the dynamics of the internal atomic variable
$s_z$ satisfies
the following equation:
\be
\dot{s}_z = \pm 2 \sqrt{s_z (s_z^2 - Ns_z - S^2 )+ C} \ \cos x  \ ,
\label{3.3}
\ee
where $C$ is an integration constant, and $x(\tau )$ is a solution of the
pendulum equation mentioned above. The equation
(\ref{3.3})
can be integrated in terms of elliptic Jacobian functions with a solution
which converges to the well-known Jaynes-Cummings semiclassical solution
\cite{JC}
in the limit  $x$ = const.

Out of resonance, at $\delta \neq 0$, the system
(\ref{3.2})
exhibits chaotic dynamics. In order to clarify the origin of  chaos,
consider Eqs.
(\ref{3.2})
in the limit of large number of excitation $N$ and large detunings $\delta$
comparing to $S^2$. Taking into account that the normalized Rabi
frequency is of the order of $\sim \sqrt{N} > 1$ and is much more larger
than the frequency of small translational oscillations, $\sqrt{\alpha} \ll
1$, the equations for the fast atom-field oscillations are reduced to the
Bloch-like form
\begin{equation}
\begin{array}{lll}
\dot{u} = \delta v \ ,\\
\dot{v} = -\delta u + 4Ns_z \cos x \ ,\\
\dot{s}_z = - v \cos x \ ,
\end{array}
\label{3.4}
\end{equation}
where the function $\cos x$ may be considered approximately as a constant
$c$ over a period of time of many Rabi oscillations. In this approximation,
the quantity
\be
u^2 + v^2 + N(2s_z )^2 = R
\label{3.5}
\ee
plays a role of the length of Bloch vector, and the general solution of
the Bloch-like equations
(\ref{3.4})
can be found
\begin{multline*}
u = u(0) \left[ N \left( \frac{2c}{\Omega_N } \right)^2 +
\left( \frac{\delta}{\Omega_N } \right)^2 \cos \Omega_N \tau \right]+\\
+ \frac{\delta}{\Omega_N} v(0) \sin\Omega_N \tau +
\frac{4N\delta c}{\Omega_N^2} s_z (0) (1-\cos\Omega_N \tau) \ ,
\end{multline*}
\begin{multline}
v = -
\frac{\delta}{\Omega_N} u(0) \sin \Omega_N \tau + v(0)
\cos\Omega_N \tau +\\
+\frac{4N c}{\Omega_N} s_z (0) \sin\Omega_N \tau \ ,
\label{3.6}
\end{multline}
\begin{multline*}
s_z = u(0) \frac{c\delta}{\Omega_N^2} (1-\cos\Omega_N \tau ) -
\frac{c}{\Omega_N} v(0)
\sin\Omega_N \tau  +\\
  +s_z(0) \left[ \left( \frac{\delta}{\Omega_N} \right)^2 + N \left(
\frac{2c}{\Omega_N} \right)^2
\cos\Omega_N \tau \right] \ ,
\end{multline*}
where the quantity
\be
\Omega_N = \sqrt{\delta^2 + (2c)^2 N}
\label{3.7}
\ee
is similar to the Rabi frequency.

Since the function $\cos x$ varies in time   slowly comparing to the
fast oscillating $u, v$, and $z$, the atom-field variable $u$ may be
considered approximately as a spatially independent
frequency- and amplitude-modulated signal that parametrically excites
the translational motion:
\be
\ddot{x} + \alpha u(\tau ) \sin x = 0 \ ,
\label{3.8}
\ee
that follows from the first two equations of the system
(\ref{3.2}).
The modulation has especially simple form
for initial conditions $u(0) = v(0) = 0$ and $s_z (0) = \mid S\mid $,
that corresponds to
the atom  prepared initially   in the upper state
while the field may be initially at any  state, and $c=1$:
\be
u(\tau ) = \frac{4N\delta \mid S \mid }{\Omega_N^2} (1-\cos \Omega_N \tau ) \ .
\label{3.9}
\ee
The Eq. (\ref{3.8}) is derived from the following
effective classical Hamiltonian
\be
{\cal H} = \frac{1}{2} \dot{x}^2 -\omega^2 \cos x + \omega^2
\cos\Omega_N \tau \cdot \cos x =
{\cal H}_0 + V \ ,
\label{3.10}
\ee
where ${\cal H}_0$ is the unperturbed Hamiltonian of a free pendulum
with the following frequency of small oscillations:
\be
\omega = \frac{2}{\Omega_N} \sqrt{\alpha N\mid \delta \mid \ \mid S \mid } \ .
\label{3.11}
\ee
Rewriting the perturbation $V = \omega^2 \cos \Omega_N \tau\cdot\cos x$
in the form
\be
V = \frac{\omega^2}{2} \left [
\cos (x+\Omega_N \tau ) + \cos (x-\Omega_N \tau )\right ] \ ,
\label{3.12}
\ee
one may consider
(\ref{3.10})
as the Hamiltonian of a particle moving in the field of three plane waves
in a frame moving with the phase velocity of the first wave,
while the phase velocities of the second and third waves
are  $\Omega_N$ and $-\Omega_N$, respectively.

As it follows from the general theory of  perturbed motion of Hamiltonian
systems with 3/2 degrees of freedom \cite{ZS91}, the Hamiltonian (\ref{3.10})
induces chaotic dynamics in the so-called stochastic layer that appears
due to the separatrix splitting. Let us consider the motion
in the neighborhood of  unperturbed separatrix of the pendulum
(\ref{3.10}).
Consider the
Poincar\'{e}-Melnikov integral
\begin{multline}
\Delta E = \int_{-\infty}^{\infty} \{ {\cal H}_0 , V \} d \tau=\\
= \omega^2 \int_{-\infty}^{\infty} \dot{x}
\sin (x- \Omega_N \tau ) d\tau \ ,
\label{3.13}
\end{multline}
where $\{ {\cal H}_0 ,V \}$
is the Poisson bracket. This integral describes changes of the atomic
translational energy at the separatrix  ${\cal H}_0 = E_s =
\omega^2$. To estimate
(\ref{3.13})
for the dynamics near the separatrix,
one can use for $x$ and $\dot{x}$ their known unperturbed
separatrix solutions
\begin{equation}
\begin{gathered}
x_s = 4 \ {\rm arctan} \exp [\pm\omega (\tau -\tau_n )]\ , \\
\dot{x}_s = \pm \frac{2\omega}{\cosh [\omega (\tau -\tau_n )] }\ ,
\end{gathered}
\label{3.14}
\end{equation}
where $\tau_n$ is introduced as an initial condition. Using the solutions
(\ref{3.14}),
we get
\be
\Delta E_n = \pm 2 \omega^2
\int_{-\infty}^{\infty}
\frac{dt}{\cosh t} \sin \left( x - \frac{\Omega_N}{\omega} t-\phi_n \right) ,
\label{3.15}
\ee
where the new time $t=\omega (\tau -\tau_n )$ and phase $\phi_n =
\Omega_N \tau_n$ were introduced. The integral
(\ref{3.15})
has been calculated to give
\begin{eqnarray}
\Delta E_n = \Delta E_s  \sin\phi_n \ ,\\
\Delta E_s = 2\pi \ \Omega_N^2 \frac{\exp (\pi\Omega_N /2\omega )}{\sinh
(\pi \Omega_N /\omega )} \ .
\label{3.16}
\end{eqnarray}
The oscillating function $\Delta E_n$ has simple zeroes that implies
transversal intersections of stable and unstable heteroclinic manifolds of
saddle points known as a complicated heteroclinic structure. On the
basis of
general properties of motion near the separatrix, the separatrix map can be
introduced \cite{ZS91}.
\begin{eqnarray}
E_{n+1} = E_n + \Delta E_s  \sin\phi_n \ ,\\
\phi_{n+1} = \phi_n + \frac{\Omega_N}{\omega} \ln
\frac{32 E_s}{|E_{n+1} -E_s |}\ .
\label{3.17}
\end{eqnarray}
The condition
\begin{multline}
K = \max \left| \frac{\delta\phi_{n+1}}{\delta\phi_n} - 1 \right|
\simeq\\
\simeq \frac{2\pi\Omega_N^3}{\omega} \
\frac{\exp (\pi\Omega_N /2\omega ) |\sin\phi_n |}{
|E_{n+1} - E_s | \sinh (\pi\Omega_N /\omega )} \gtrsim \ 1
\label{3.18}
\end{multline}
defines the stochastic layer width which can be estimated in the case of
the large
parameter $\Omega_N /\omega \gg 1$ (see the respective estimations with
real atoms in the concluding section) as follows:
\be
\delta E_s \equiv |E_{n+1} - E_s | \lesssim
\frac{8\pi\Omega_N^3}{\omega} \exp
\left( - \frac{\pi\Omega_N}{2\omega} \right) .
\label{3.19}
\ee
The dimensionless width of the stochastic layer is finally given by
\be
\frac{\delta E_s}{\omega^2}  \simeq 8\pi
\left(  \frac{\Omega_N}{\omega} \right)^3
\exp \left( -\frac{\pi\Omega_N}{2\omega} \right)\ ,
\label{3.20}
\ee
where the large parameter
\be
\frac{\Omega_N}{\omega} =
\frac{\delta^2 + 4N}{2\sqrt{\alpha N\mid \delta \mid  \ \mid S \mid } }
\label{3.21}
\ee
under the conditions $N,\delta \gg \mid S \mid  \simeq 1$ and
$\delta^2 \ll 4 N$ is estimated as
\be
\frac{\Omega_N}{\omega} \simeq
2 \sqrt{\frac{N}{\alpha \mid \delta \mid }}\ .
\label{3.22}
\ee
For the considered case the width of the stochastic layer  of the reduced
atom-field dynamics is exponentially small in (\ref{3.20}),
multiplied by a large parameter. Due to (\ref{3.22}) the final width depends
on the control parameters $N$, $\alpha$ and $\delta$, 
and the formula (\ref{3.20}) is useful in estimating the ranges of the 
control parameters where one may expect chaotic motion.  

The estimation (\ref{3.20}) provides the lower bound for the 
the width of the stochastic layer that appears due to the simplest 
harmonic modulation  (\ref{3.9}). Small changes in energy produce 
comparatively small changes in frequency of oscillations. Nearby the bottom of 
potential wells and high over potential hills (where the energy is much less 
and much greater than $E_s$), small changes in frequency give rise to respectively 
small changes in phase during the period of oscillations. Nearby the 
unperturbed separatrix, where the period goes to infinity, even small 
changes in frequency lead to dramatic changes in phase. This is 
the reason of exponential instability of the parametric oscillator (\ref{3.8}) 
and (\ref{3.9}) which models chaotic motion of the atom moving through 
a periodic standing wave.

\section{Lyapunov exponents and Poincar\'{e} sections}

In this section, we present numerical simulations with the basic set of
equations (\ref{2.3}) and the integrals of motion (\ref{2.4}) and 
(\ref{2.5}) with $S^2 = 3/4$
(the actual value of $S^2$ has no importance since it always can be
renormalized to 1)
and $N = 10$. The system
(\ref{2.3})
has two control parameters, the normalized detuning $\delta$ between the
atomic transition and cavity frequencies and the normalized recoil frequency
$\alpha$. As it will be estimated in the concluding section,
$\alpha$ is in the range from $10^{-5} $ to $10^{-2}$ for real atoms.
We choose $\alpha = 10^{-3}$ in  simulations throughout the paper.

\begin{figure}[htb]
\includegraphics[width=0.4\textwidth,clip]{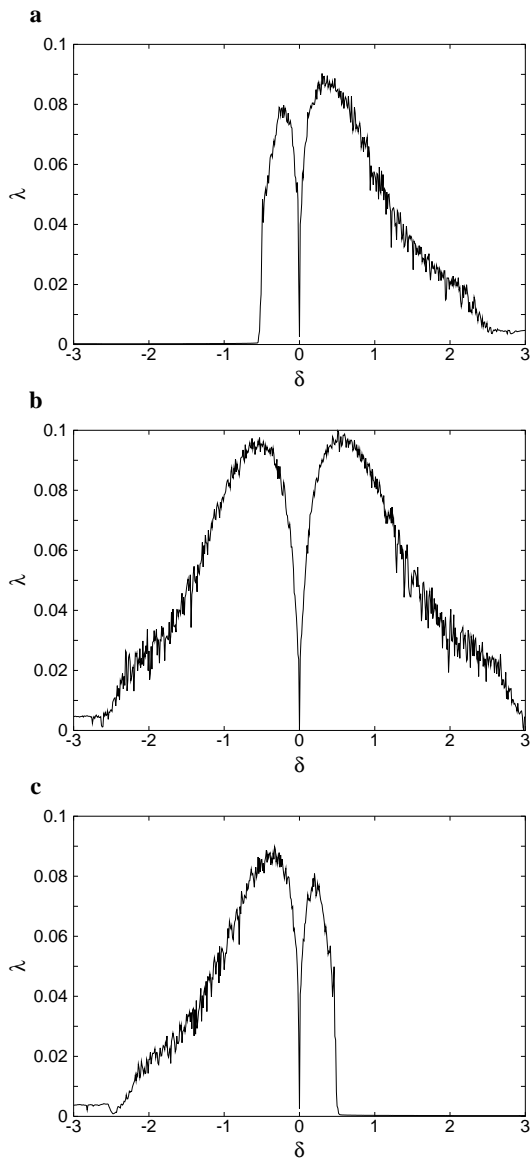}
\caption{\label{fig1}The maximal Lyapunov exponent $\lambda$ in units of the maximal
atom-field coupling rate $\Omega_0$ versus the atom-field detuning $\delta$
in units $\Omega_0$. (a) $s_z (0) = -0.863$. (b) $s_z (0) = 0$.
(c) $s_z (0) = 0.863$.}
\end{figure}
\begin{figure*}[htb]
\hspace{0.05\textwidth} {\large\bf a} \hspace{0.324\textwidth} {\large\bf c}
\hfill $\mathstrut$\\
\includegraphics[width=0.324\textwidth,clip]{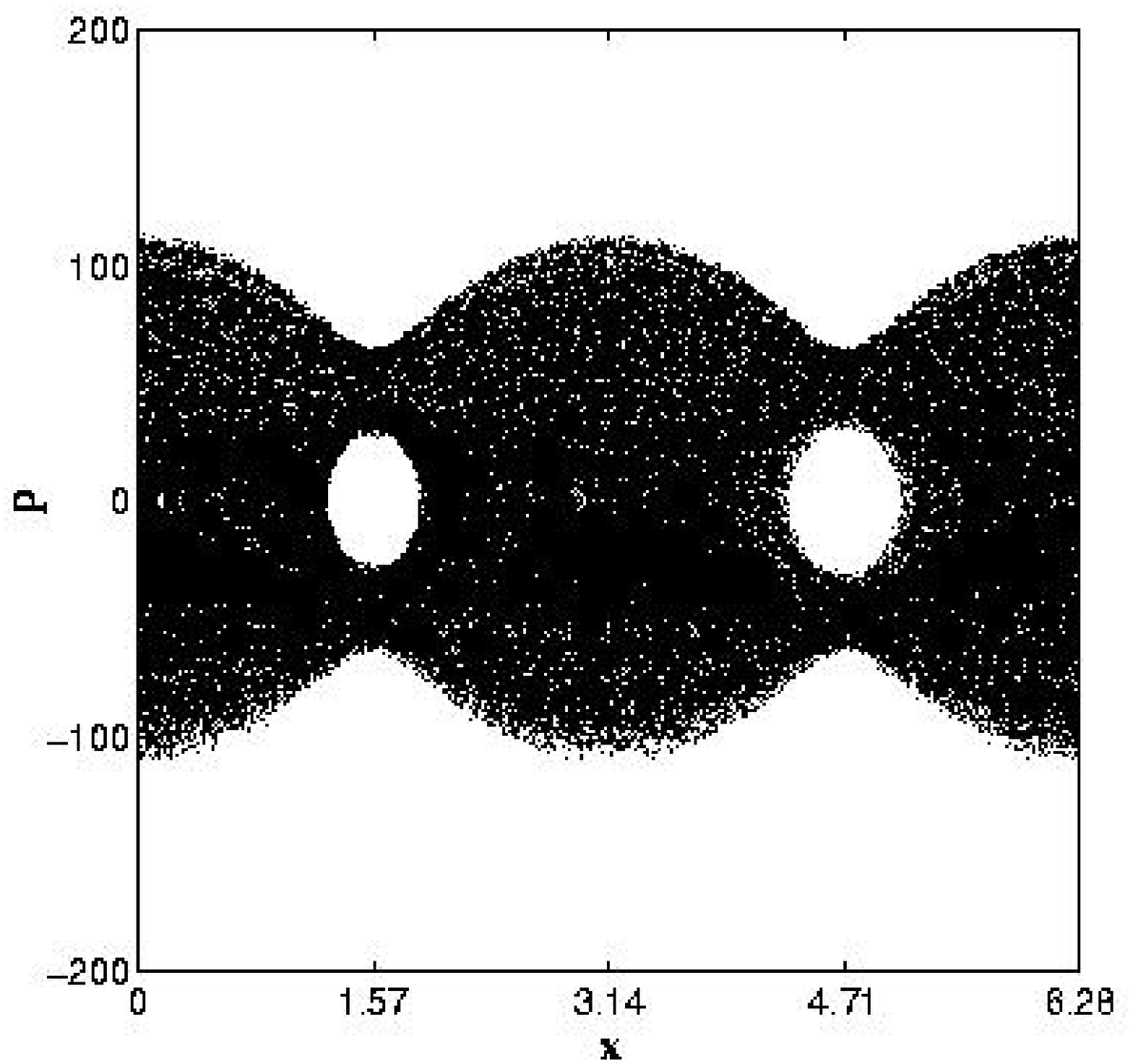}
\includegraphics[width=0.576\textwidth,clip]{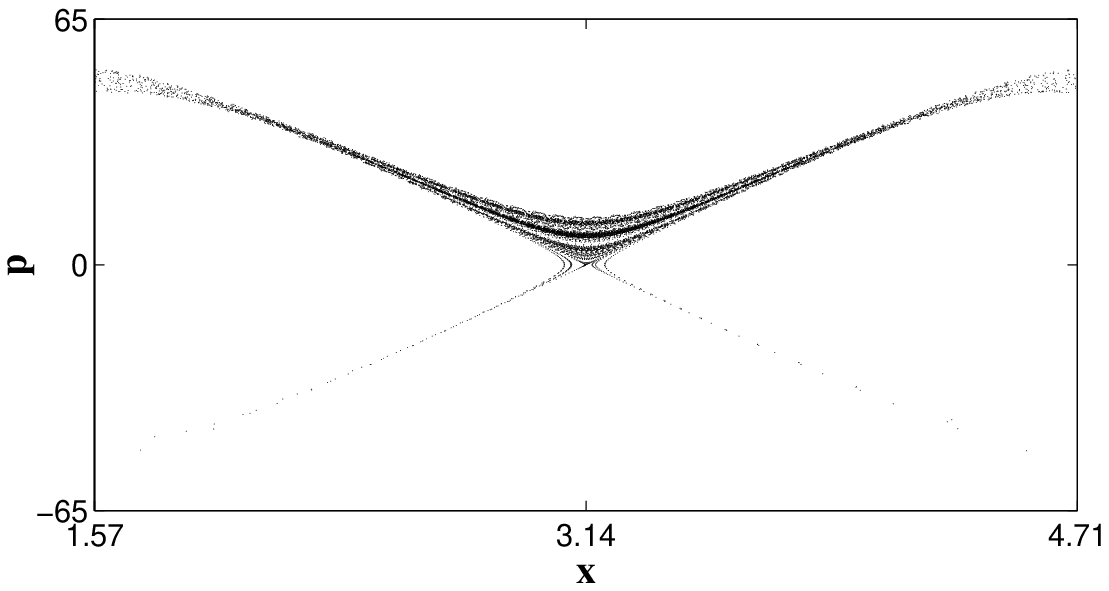}\\[1mm]
\hspace{0.05\textwidth} {\large\bf b} \hspace{0.324\textwidth} {\large\bf d}
\hfill $\mathstrut$\\
\includegraphics[width=0.324\textwidth,clip]{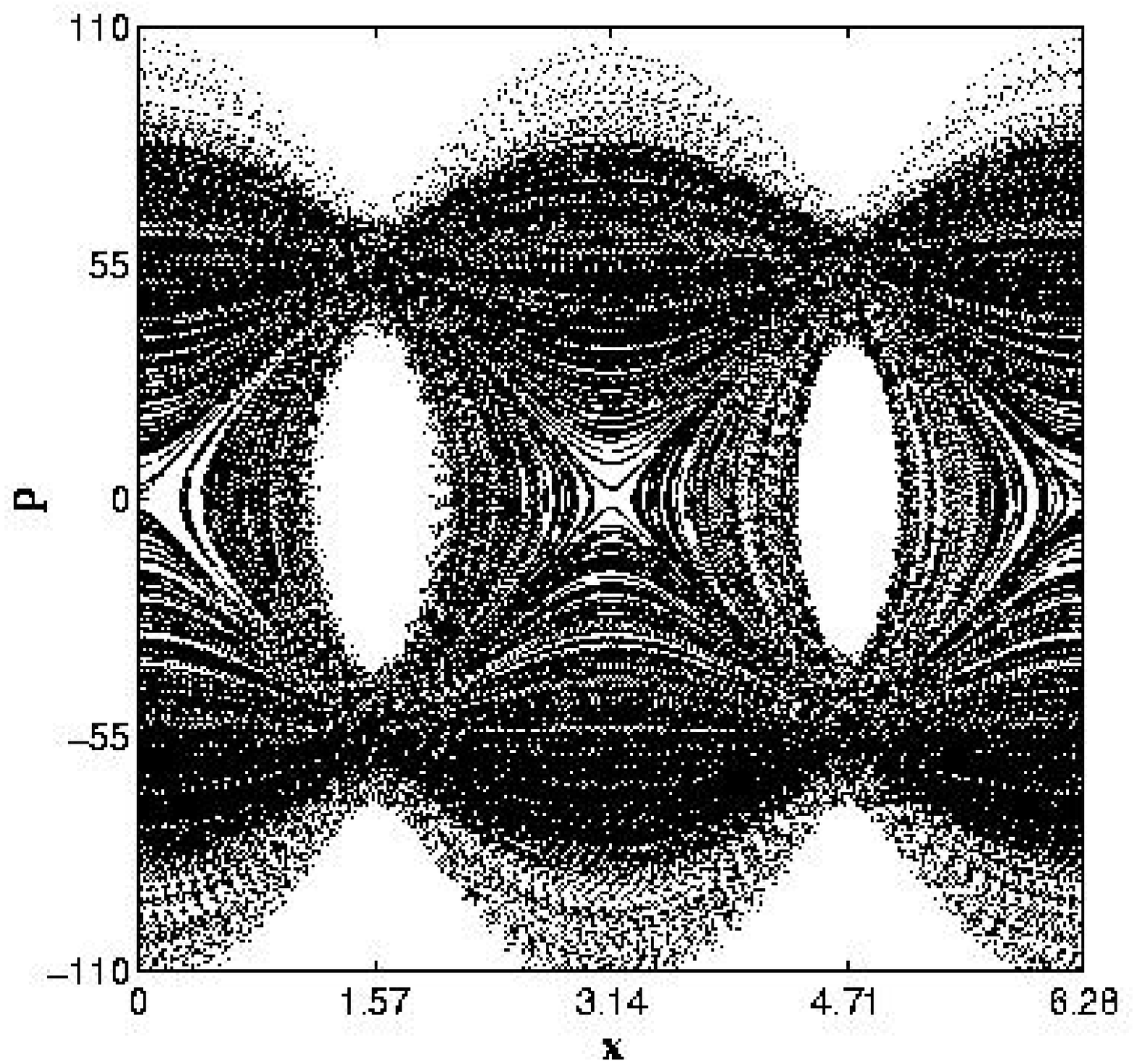}
\includegraphics[width=0.576\textwidth,clip]{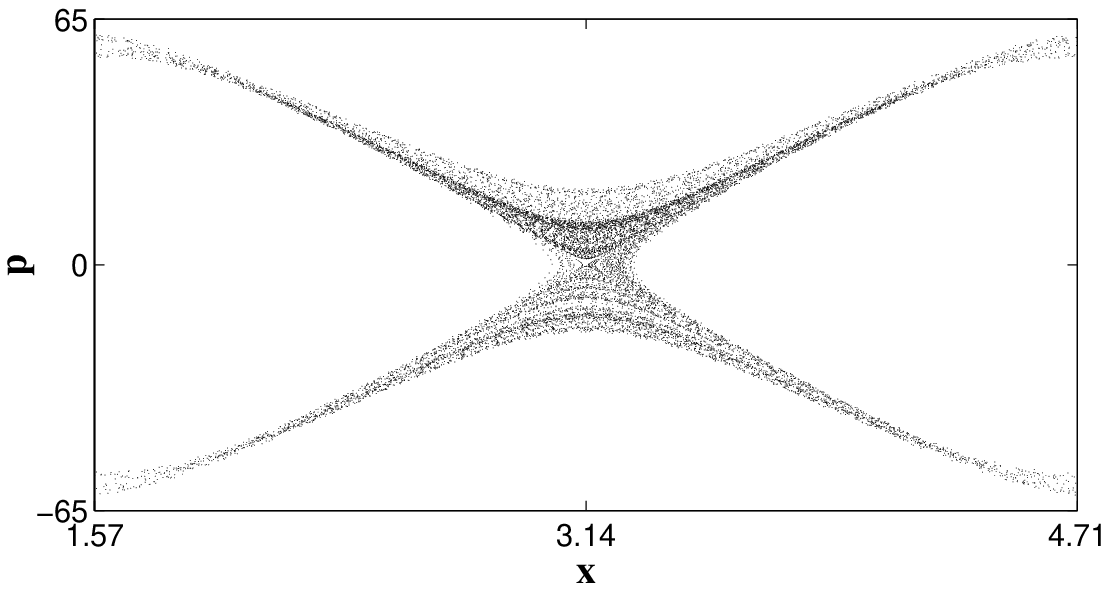}\\
\caption{\label{fig2}Projections of the Poincar\'{e} sections on the plane of the atomic momentum
$p$ in units $\hbar k_f$ and the position in units of $k_f^{-1}$.
(a) $s_z (0) =-0.8660254$ and $\delta = 1.2$. (b) $s_z (0) = - 0.863$ and
$\delta = 1.2$.
(c) $s_z = -0.8660254$, $\delta = 1.92$, and $p(0) = 2.1$ in all the above
fragments. (d) $s_z = -0.863$, $\delta = 1.92$, and $p(0) = 20$. $x, p$ are dimensionless.}
\end{figure*}

The detuning $\delta$, as it was shown in \cite{PK01,PS01},
is the crucial parameter in transition to chaos in the atom-field system
with the center-of-mass motion. It is obvious from the set (\ref{3.2}),
which is equivalent to the basic one (\ref{2.3}),
that at exact resonance, $\delta = 0$, the motion is regular. At large
detunings, $\delta \gg N$, the motion is expected to be quasiregular
since the nonlinear term in the fourth equation of the set (\ref{3.2})
is small, compared to the linear term of the same
equation. With far-detuned light, one does not expect pronounced atomic Rabi
oscillations. In order to find the range of the detunings, where the motion
is expected to be chaotic, we compute the dependence of the maximal
Lyapunov exponent $\lambda$ on the detuning $\delta$.

Lyapunov exponents characterize the behavior of close trajectories in phase 
space. Consider a trajectory, some sequence of time instants $\tau_0$, $\tau_1$,
$\tau_2$, ... with equal intervals $\Delta \tau$, and a ball around the initial 
point of the trajectory. Lyapunov numbers $\Lambda_{j}$ ($j=1,...,M$;
M is a number of variables) show per-interval   $\Delta \tau$ changes of the
axes of the ``ellipsoid'' of the deformed ball \cite{ASY}. In our case
$M=6$ after exclusion of $s_z$.
The $k$-th Lyapunov exponent is defined  as $\lambda_{k}=\ln \Lambda_{k} 
$.
Typically  $\Lambda_{k}$ depends on time and  $\lambda_{k}$ should be
replaced by their mean values \cite{ASY}. In Hamiltonian systems, due to
the phase volume conservation, $\Lambda_{1} \cdot \Lambda_2 \cdot ... \cdot
\Lambda_{M} = 1$ and
$\lambda_{1}+...+\lambda_{M} = 0$. For integrable system all
$\Lambda_{k}$
are pure imaginary and they make pairs: $\lambda_{2}=-\lambda_{1}$,
$\lambda_{4}=-\lambda_{3}$, $\lambda_{6}=-\lambda_{5}$, since 
the number $M$ of equations is even. This result follows from the
so-called Liouville-Arnold theorem \cite{A88}. In our case, for
the reduced system of 6 variables ($p,x;a_x,a_y;s_x,s_y$) and two
constrains (integrals of motion), we have two imaginary pairs, say
$\lambda_{1,2}= \pm i \sigma_1$,  $\lambda_{3,4}= \pm i \sigma_2$  
($\sigma_{1,2}$ real),
and $\lambda_{5}$, $\lambda_{6}$ that satisfy the condition 
$\lambda_{5}+\lambda_{6} = 0$, and therefore 
$\Lambda_{5} \cdot \Lambda_{6} = 1$, i.e. $\Lambda_{6} = 1/\Lambda_{5}$.
The chaos means that $\lambda_{5,6}$ are real \cite{LL,ASY}.
If, say $\Lambda_{5} < 1$ ($\lambda_{5} < 0$), then  $\Lambda_{6} > 1$ 
($\lambda_{6} > 0$) and $\lambda_{6}$ is called
maximal Lyapunov exponent. It has a nice physical meaning: the 
maximal Lyapunov exponent measures a rate of the separation of
initially close trajectories, and typically for the practical goal,
the mean value $\lambda = \overline {\lambda_6}$ over time
is used. To compute $\lambda$, we use the standard algorithm 
\cite{LL}
\begin{equation}
\lambda=\lim_{n{\rightarrow}\infty}
\frac {1}{n} \sum_{k=1}^{n} \ln \frac{\Delta (\tau_{k})}{\Delta (\tau_{k-1})},
\label{4.1}
\end{equation}
where $\Delta (\tau_{k-1})$ is a distance between two close trajectories at
time $\tau_{k-1}$, and the value $\Delta (\tau_{k})/\Delta (\tau_{k-1})$ shows the
level of separations of the trajectories during the interval
$(\tau_{k-1},\tau_{k})$. 

In the  case that separation doesn't go exponentially, $\lambda = 0$.
This happens when $\delta = 0$ since the system becomes integrable,
and the exponential separation disappears.

The corresponding results of computing maximal
Lyapunov exponent $\lambda$  are presented in FIG.~\ref{fig1} with
three different initial values of the atomic population inversion
$s_z (0) = - 0.863$; 0; 0.863, respectively. The other initial conditions are the
following: $a_x (0) = s_x (0) = 0$, $x(0) = 3.14$, $p(0) = 2$, and
$a_y (0)$ and $s_y (0)$ are found from the Eqs.(\ref{2.4}) and (\ref{2.5})
with given $S = \sqrt{3} /2$ and $N =10$. The value $s_z (0) = -0.863$
corresponds to the atom initially prepared closely to its ground state
for which $s_z = - \sqrt{3} /2$. Note that the unusual amplitude values of
the atomic population inversion we have are the result of the
chosen normalization
$S^2 = 3/4$. The atom with $s_z (0) = 0.863$ is prepared closely to
its excited state. In both the cases, the initial components of the
transition electric dipole moment, $s_x (0)$ and $s_y (0)$, are almost zero
with $|s_z (0)| = 0.863$. The atom with $s_z (0) = 0$ has a maximal electric
dipole moment. 

As it was expected, at exact resonance $(\delta =0)$, the maximal 
Lyapunov exponent is exactly equal zero in all the cases assuming 
a regular motion. As it follows from the results of previous section, 
a stochastic layer appears with infinitesimally small values of detuning, 
but its width decreases fast with increasing $\delta$ (see (\ref{3.20}) 
and (\ref{3.21})). We find $\lambda \simeq 0$ with $|\delta | \gtrsim 3$. 
In physical terms, it means that at exact resonance an atom will periodically 
exchange excitation with the field, whereas far off resonance its internal 
states will not (almost) be affected by the field. This interplay results in 
a maximum of the $\lambda (\delta)$ dependence with almost the same 
maximal values for $s_z (0) = \pm 0.863$ and 0. 
The results, however, are different in the
range $-3 \lesssim \delta \lesssim 3$
with different initial values of the population inversion $s_z$. It is easy
to understand why it is. As it follows from the Bloch-like solution
(\ref{3.6}) for $s_z$, the atom starting, say, in its ground state,
$s_z (0) =- \sqrt{3}/2$ and $u(0) = v(0) = 0$, could reach the upper state
$s_z = \sqrt{3}/2$ only with
$\delta = 0$. The same is valid with the other initial values of $s_z (0)$:
the atom starting with $s_z (0) = \sqrt{3}/2$ (or with $s_z (0) = 0$) will
not reach $s_z = - \sqrt{3}/2$ (or $s_z = \pm \sqrt{3}/2)$ except for the
case of exact resonance, $\delta = 0$. Thus, the dependencies $\lambda
(\delta )$ are different in the range
$-3 \lesssim \delta \lesssim 3$
with different values of $s_z (0)$ in spite of the fact that the maximal
Lyapunov exponent is computed over the rather long trajectory.

The model Hamiltonian (\ref{2.1})
can be easily generalized to an ensemble of indistinguishable two-level
atoms. In the semiclassical approximation, we have not observed any
pronounced differences in the strength of chaos (that is characterized by
the values of $\lambda $) with different initial internal atomic states.
Interatomic quantum correlations, which occur through the mediation of the
field generated by the atomic ensemble, have been shown in \cite{PK00}
(where the model with hot moving atoms but without recoil has been considered)
to play a significant role in the atom-field dynamics. Much more strong
chaos has been numerically found \cite{PK00}
in the vacuum Rabi oscillations with atoms initially prepared in the so-called
superfluorescent state (with all the atoms to be uncorrelated initially and
occupying their excited states) than with atoms initially prepared in the
superradiant state (with initially strongly correlated atoms having a
macroscopic electric dipole moment).

We numerically construct single-trajectory Poincar\'{e} sections
of motion in the system (\ref{2.3})
with three degrees of freedom and project them on the plane of the atomic
external variables $(x,p)$. FIG.~\ref{fig2}
presents these sections with the atom initially prepared close to the
ground state and with two different values of the atom-field detuning
$\delta = 1.2$ (see (a) and (b) fragments) and $\delta = 1.92$
(see (c) and (d) fragments).
In the latter case chaos is not as strong as in the
first case  (see FIG.~\ref{fig1}a).
A fairly regular web structure, that is seen in the fragment (b) computed
over a comparatively short integration time, breaks down with increasing
integration time (the fragment (a)). For comparison, we present in
FIG.~\ref{fig2}c the Poincar\'{e} section computed under the same conditions as in
FIG.~\ref{fig2}a but
with $\delta = 1.92$ (an additional trajectory with $p(0) = 0.2$ is plotted
in the fragment (c)). FIG.~\ref{fig2}d demonstrates the Poincar\'{e} section
at $\delta = 1.92$ with an increased initial momentum $p(0) = 20$.

\section{Statistical properties of the atom-photon interaction}

\begin{figure}[htb]
\hspace{0.05\textwidth} {\large\bf a} \hfill $\mathstrut$\\
\includegraphics[width=0.45\textwidth,clip]{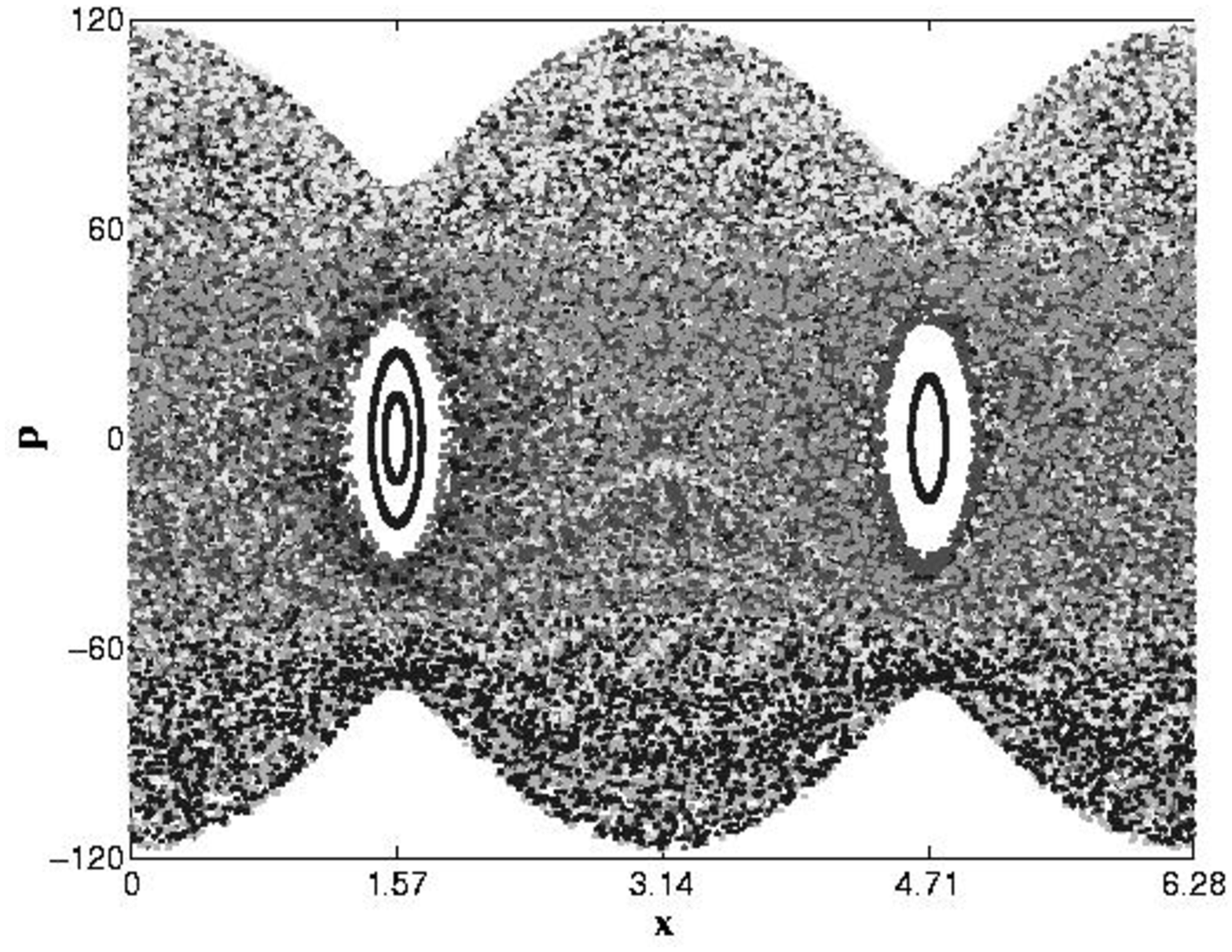}\\[2mm]
\hspace{0.05\textwidth} {\large\bf b} \hfill $\mathstrut$\\
\includegraphics[width=0.45\textwidth,clip]{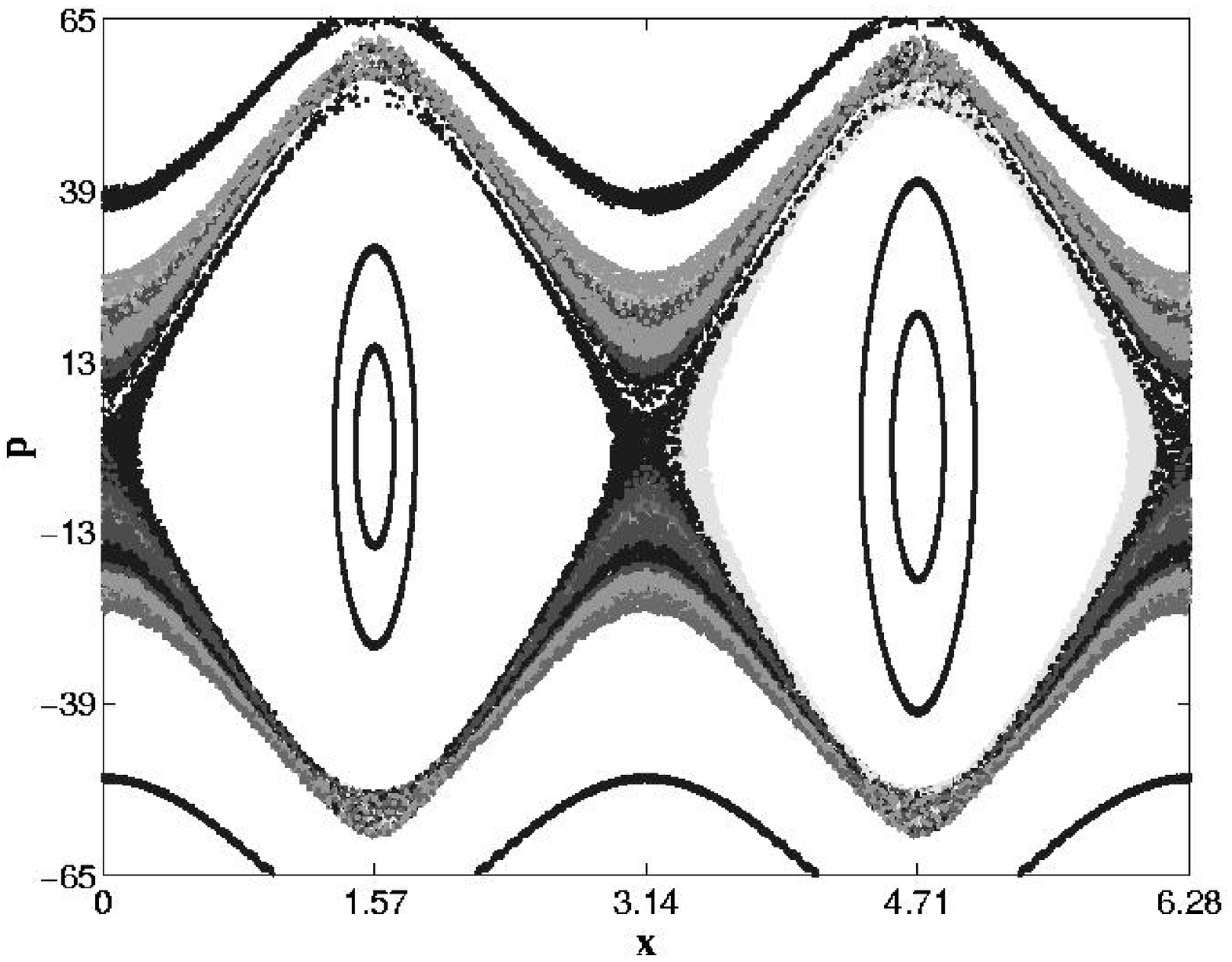}
\caption{\label{fig3}The same as in FIG.~\ref{fig2} with $s_z (0) =-0.863$ and $p(0)=1.$
Change of density appears after (a) $\Delta \tau =75,000$
and $\delta = 1.2$ and (b) $\Delta \tau =49,000$ and $\delta = 1.92$. $x, p$ are dimensionless.}
\end{figure}
\begin{figure}[!htb]
\hspace{0.05\textwidth} {\large\bf a} \hfill $\mathstrut$\\
\includegraphics[width=0.4\textwidth,height=0.25\textheight,clip]{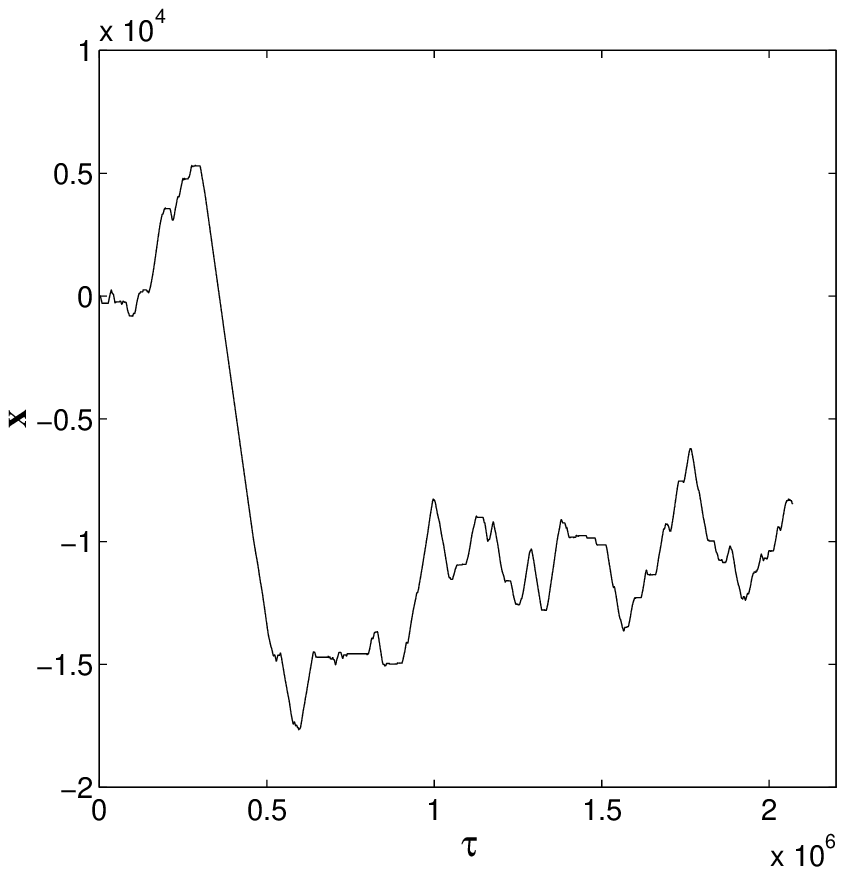}\\[1mm]
\hspace{0.05\textwidth} {\large\bf b} \hfill $\mathstrut$\\
\includegraphics[width=0.4\textwidth,height=0.25\textheight,clip]{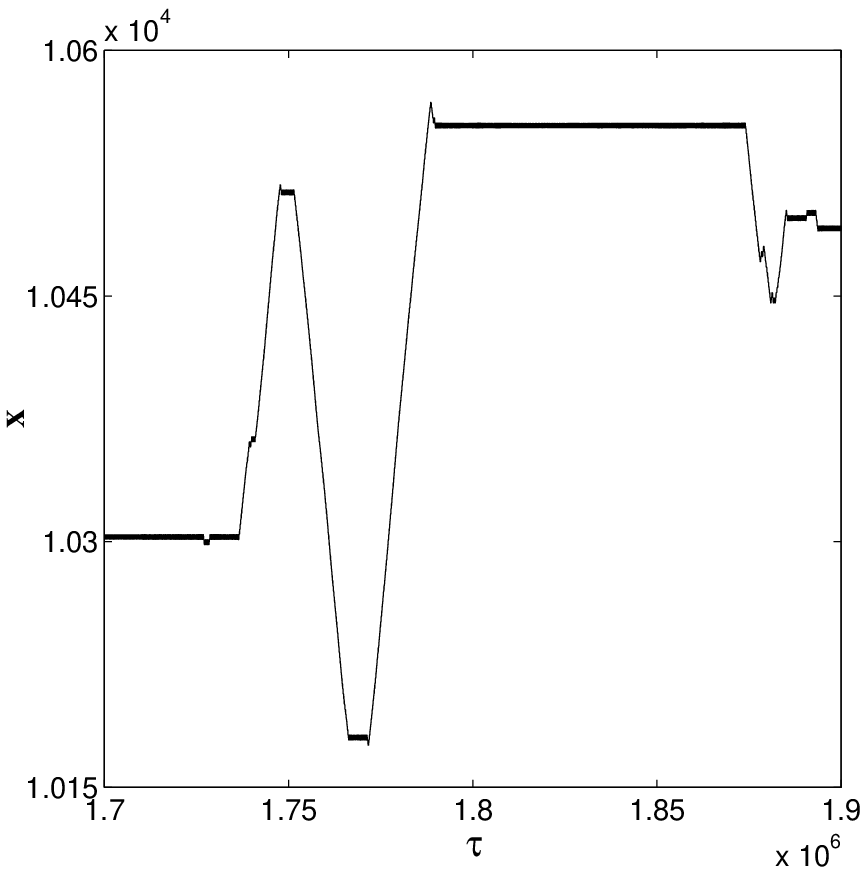}\\[1mm]
\hspace{0.05\textwidth} {\large\bf c} \hfill $\mathstrut$\\
\includegraphics[width=0.4\textwidth,height=0.25\textheight,clip]{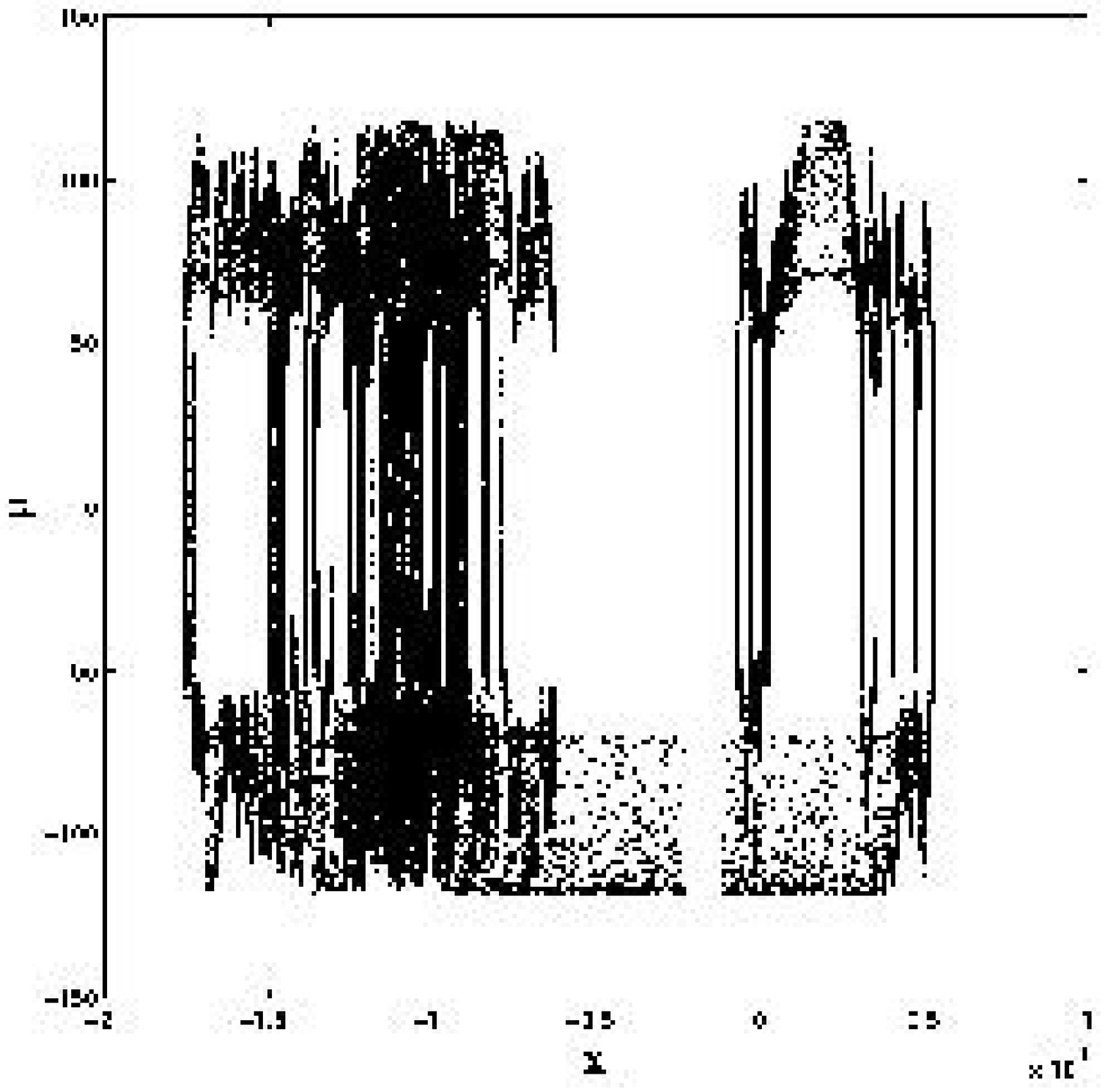}
\caption{\label{fig4}L\'{e}vy flights of an atom in a cavity. (a) A long ballistic flight at
$\delta = 1.2$ is evident. (b) Ballistic flights intermittent with stagnation
phases of motion at $\delta = 1.92$ are shown. (c) Two different types of the
L\'{e}vy flights in the plane $x - p$. Time is in units of
$\Omega_0^{-1}$. $x, p, \tau$ are dimensionless. }
\end{figure}

In this section a detailed analysis of the atom and photon chaotic dynamics
will be considered. A sensitive control parameter is $\delta = (\omega_f -
\omega_a )/\Omega_0$, detuning of the field and atom frequencies. For the
sake of convenience we specify two values of $\delta$: 1.2 and 1.92. It follows
from calculating maximal Lyapunov exponents in the previous section that
the smaller is $\delta$ (in the range $0.5 \lesssim   \mid \delta \mid 
\lesssim   3$), the stronger is mixing
and chaos, and one can expect that the case with $\delta = 1.92$ is, being
chaotic (but not with the atom prepared in the upper  state or close to it),
more intermittent than the case $\delta = 1.2$. This property of the
atom-photon dynamics will be quantitatively characterized below.

The difference of a trajectory projection on the $x - p$  plane
$(a_x = s_x = 0$) is evident from FIG.~\ref{fig3} where the density modulation has
been used: a change of each density  appears after $\Delta \tau$ points of the
mapping
the trajectories ($\Delta \tau = 75,000$ for (a) and $\Delta \tau = 49,000$ for
(b)). The narrow strips of the same density  in FIG.~\ref{fig3}b indicate a long stay of
atom in the corresponding part of the $x - p$ plane with oscillations in the
potential well and a small change of the amplitude of the oscillations.
In contrast to this pattern, the distribution of densities  in FIG.~\ref{fig3}a is more
uniform manifesting much better mixing, although some traces of the
intermittency persist.

The difference between $\delta = 1.2$ and $\delta = 1.92$ is also evident from
FIG.~\ref{fig4} where a dependence $x = x(\tau)$ is shown. The intermittent case
($\delta = 1.92$) has very long ``flights'' known also as
L\'{e}vy flights \cite{SZK,Z98}. There are two types of flights in FIG.~\ref{fig4}b.
One  category of
flights corresponds to the almost linear dependence of $x = x(\tau)$, while the
other one corresponds to the stagnation of the trajectory near some value of
$x$. FIG.~\ref{fig4}c shows the flight in the $x - p$ plane where the ballistic
dynamics coexists or alternate stagnations. Both categories of flights are
well understood from FIG.~\ref{fig2}c,d: ballistic dynamics along $x$
in FIG.~\ref{fig2}c is
responsible for the linear dependence of $x = x(\tau)$, while the trajectory
can stay very long near the saddle points as in FIG.~\ref{fig2}d (the dark area near
a saddle point). The case of $\delta = 1.2$ in FIG.~\ref{fig2}a,b is very
different and flights of both categories are rare, if ever.

Just the presence of flights and intermittent behavior of the physical
variables strong fluences the statistical properties of atoms and photons.
We will use two important characteristics of the atom-photon variables:
distribution of Poincar\'{e} recurrences and moments of the atom
coordinate $x$. Consider a small phase volume $\Delta\Gamma$ and
$P(\Delta\Gamma ;\tau)$ as a probability density of a trajectory to return
first time back to $\Delta\Gamma$ at time instant $\tau \in (\tau+d\tau)$ if
initially started at $\Delta\Gamma$ at $\tau=0$. Then the density
probability to return first time to $\Delta\Gamma$ is
\be
P(\tau) = \lim_{\Delta\Gamma\rightarrow 0} \frac{1}{\Delta\Gamma}
P(\Delta\Gamma ;\tau) \,
\label{A.1}
\ee
with a normalization condition
\be
\int_0^{\infty} P(\tau) d\tau =
\lim_{\Delta\Gamma\rightarrow 0} \frac{1}{\Delta\Gamma}
\int_0^{\infty}
P(\Delta\Gamma ;\tau) d\tau = 1 \ .
\label{A.2}
\ee
The probability $P(\tau)$ does not depend on the choice of $\Delta\Gamma$
and for ``good'' chaotic mixing decays exponentially \cite{ZEN}
\be
P(\tau) = (1/h)e^{-h\tau}\,
\label{A.3}
\ee
with the mean recurrence time
\be
\tau_{\rm rec} = 1/h =
\int_0^{\infty} \tau P(\tau) d\tau \ ,
\label{A.4}
\ee
and $h$ as Kolmogorov-Sinai entropy.

\begin{figure}[htb]
\includegraphics[width=0.4\textwidth,clip]{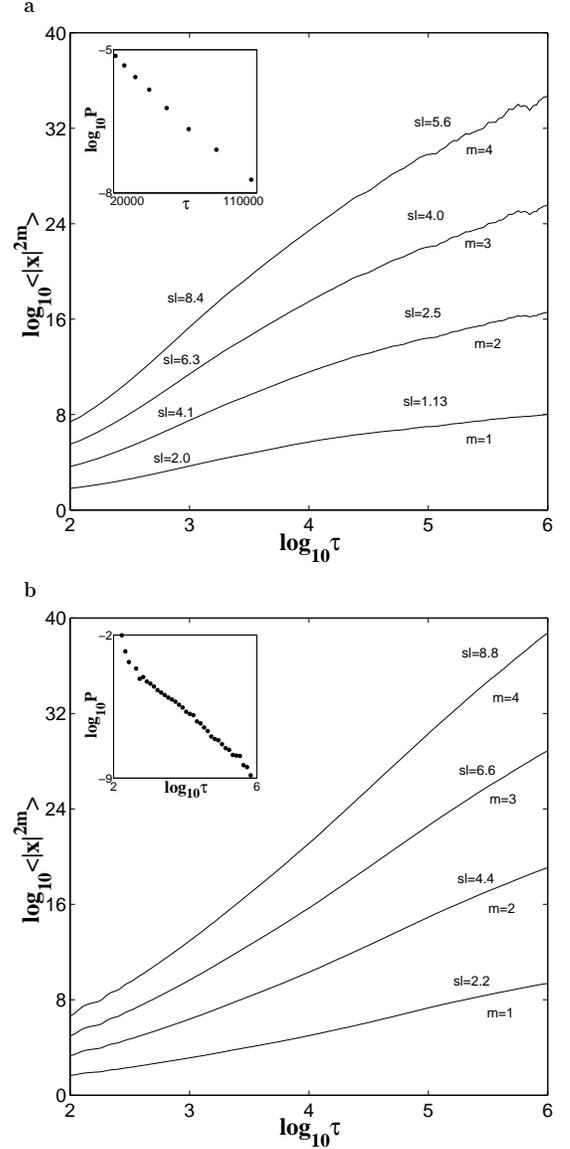}
\caption{\label{fig5}Time evolution of the $2m$th-order moments of the atom position $x$ on
a logarithmic scale with the values of slopes indicated for each $m$.
The insets show the respective distributions of the Poincar\'{e} recurrences.
(a) $\delta = 1.2$. (b) $\delta = 1.92$. $x, \tau$ are dimensionless.}
\end{figure}

The general situation is more complicated since an algebraic behavior
\be
P(\tau) \sim 1/\tau^{\gamma}\ , \quad
\tau \rightarrow \infty \ ,
\label{A.5}
\ee
is possible for large $\tau$ due to intermittent chaos. For bounded
Hamiltonian dynamics, $\tau_{\rm rec} < \infty$
(Kac lemma), and the condition $\gamma > 2$ should exist. Nevertheless,
strongly intermittent dynamics sometimes does not permit us to achieve the
limit at $\tau \rightarrow \infty$
and many different intermediate asymptotics can appear. FIG.~\ref{fig5} shows
the distribution of recurrences that is close to the exponential one as in
(\ref{A.3}) for $\delta = 1.2$ and to the algebraic one as in (\ref{A.5}) with
$\gamma \gtrsim 2$ $(\tau > 10^5$) for $\delta = 1.92.$

The difference due to intermittency also occurs for the moments
\be
\langle x^{2m} \rangle \sim \tau^{\mu (m)} \ ,
\label{A.6}
\ee
where the so-called transport exponent $\mu (m)$ varies for different
time windows. The behavior of $\langle x^{2m} \rangle$
is shown in FIG.~\ref{fig5}. For $\delta = 1.2$, the value $\mu (1)$ is close to 2
for $\tau < 10^3$ and corresponds to the ballistic dynamics. For $\tau > 10^3$
and $\delta = 1.2$, we have $\mu (1) \approx 1.13$ that corresponds to a
weak superdiffusion that is fairly close to the normal diffusion with
$\mu (1) = 1$.  A very different behavior for moments appears for $\delta
= 1.92$ where there are many long-lasting flights. For $\tau > 10^3$,
$\mu (1) \approx 2.2$ that corresponds to a superballistic transport with an
acceleration. This behavior can be explained as a result of long flights
when atoms move in the photon's field acquiring acceleration. This type of
transport is self-similar and $\mu (4) \approx 8.8 = 4 \mu (1)$.

\section{Manipulation of Atoms}

In this section we would like to make a few comments related to the
manipulation of atoms by changing different control parameters. As it was
shown in \setcounter{abcd}{5}Sec.~\Roman{abcd}, a change of $\delta$ leads to a possibility of a
sensitive control of the L\'{e}vy flights
and, as a result, to cool the atoms which have the lower chaotic
dispersion the longer the flight is. At the same time, simulations show
fast mixing on the $a_x - a_y $ and $s_x - s_y $ planes. More precisely,
spectral properties of the atomic dynamics are sensitively controlled by the
parameter $\delta$. Let us demonstrate it using a simplified analysis.

Consider $x = x(\tau)$ as the only variable that describes the dynamics
or the most essential part of the dynamics, and introduce a generation
function
\be
G(x,\tau; \nu ) = e^{i\nu [x(\tau)-x_0 ]} \ , \ \ \ \ \
x_0 \equiv x(0) \ .
\label{B.1}
\ee
Then
\be
I(\tau;x_0 ) \equiv
\int_{-\infty}^{\infty}
d\nu \ G(x,\tau; \nu ) = 2\pi\delta [ x(\tau) -x_0 ] \ .
\label{B.2}
\ee
The expression $I(\tau;x_0 )$ can be ``coarse-grained'' over $x_0$, i.e.
\begin{multline}
\langle I(\tau;x_0 )\rangle_{x_0} \equiv \frac{1}{\Delta x_0} \int dx_0\,
I(\tau, x_0)= \\
 = \frac{2\pi}{\Delta x_0} \int dx_0 \ \delta [x(\tau)-x_0 ]\ .
\label{B.3}
\end{multline}
The presence of $\delta$-function indicates recurrences to $x_0$ within an
interval $\Delta x_0$ at time instant $\tau$ within an interval $\Delta \tau_0$.
For $\tau \rightarrow\infty$ we can neglect $n$-triple recurrences from
$n \ge  2$ and leave only the first recurrences. Then
\begin{multline}
\langle I(\tau;x_0 )\rangle_{x_0} = P(\tau) =\\
=\sum_{x_0 \in \Delta x_0} {\rm const} / \left| \frac{dx(\tau)}{dx_0}
\right|_{x(\tau) = x_0}\ .
\label{B.4}
\end{multline}
The expression (\ref{B.4}) shows that for ``good'' chaotic systems
\be
|dx(\tau)/dx_0 | \sim\exp (h\tau) \ ,
\label{B.5}
\ee
and we arrive at (\ref{A.3}). For the intermittent dynamics, the sum in (\ref{B.4})
consists of two types of terms, the same as (\ref{B.5}) and the algebraic growth
\be
|dx/dx_0 | \sim \tau^{\gamma} \ ,
\label{B.6}
\ee
with the value of $\gamma$ depending on the type of intermittency. For
a fairly large $\tau$, the term (\ref{B.6}) survives and we arrive at the case
(\ref{A.5}).

From another side,
\be
\frac{\partial^{2m}}{\partial\nu^{2m} } \langle G(x,\tau; \nu
\rangle_{\nu =0} = (-1)^m \langle |x(\tau) - x_0 |^{2m} \rangle \ ,
\label{B.7}
\ee
and we obtain the moments of $x(\tau)$. This shows that the moments and their
spectral properties are coupled to the recurrences distribution through the
generating function $G(x,\tau;\nu )$ which one would expect to obtain from
experiments. When the moments are infinite, the expression (\ref{B.7}) can be
replaced by the following one
\be
\frac{\partial^{\beta}}{\partial\nu^{\beta} }
\langle G(x^{\beta},\tau; \nu^{\beta} \rangle_{\nu =0}
= {\rm const} \   \langle |x(\tau) - x_0 |^{\beta} \rangle \ ,
\label{B.8}
\ee
with an appropriate value of $\beta$ (see more discussion in \cite{SZ97}).
\begin{figure}[htb]
\hspace{0.05\textwidth} {\large\bf a} \hfill $\mathstrut$\\
\includegraphics[width=0.4\textwidth,clip]{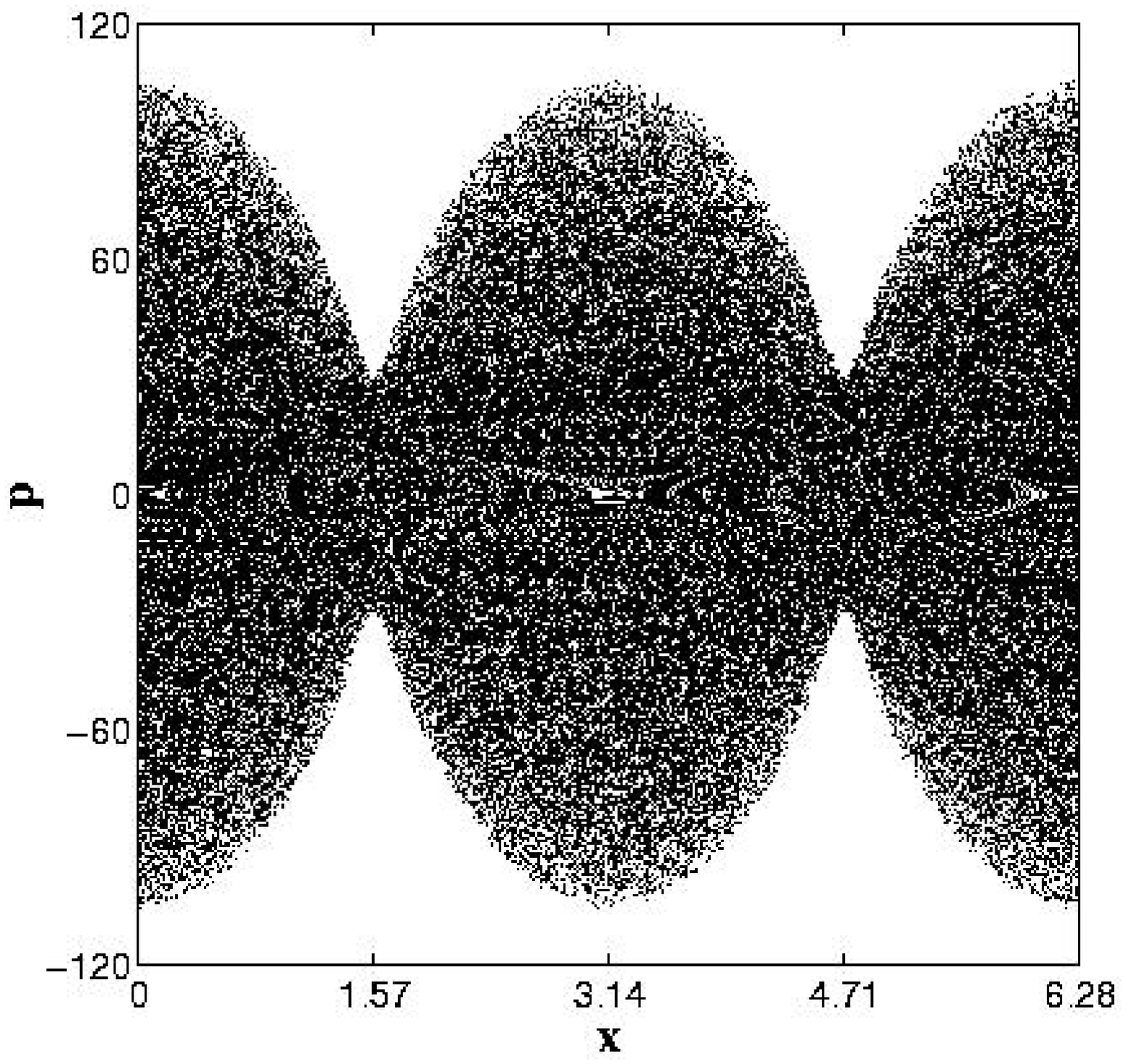}\\[1mm]
\hspace{0.05\textwidth} {\large\bf b} \hfill $\mathstrut$\\
\includegraphics[width=0.4\textwidth,clip]{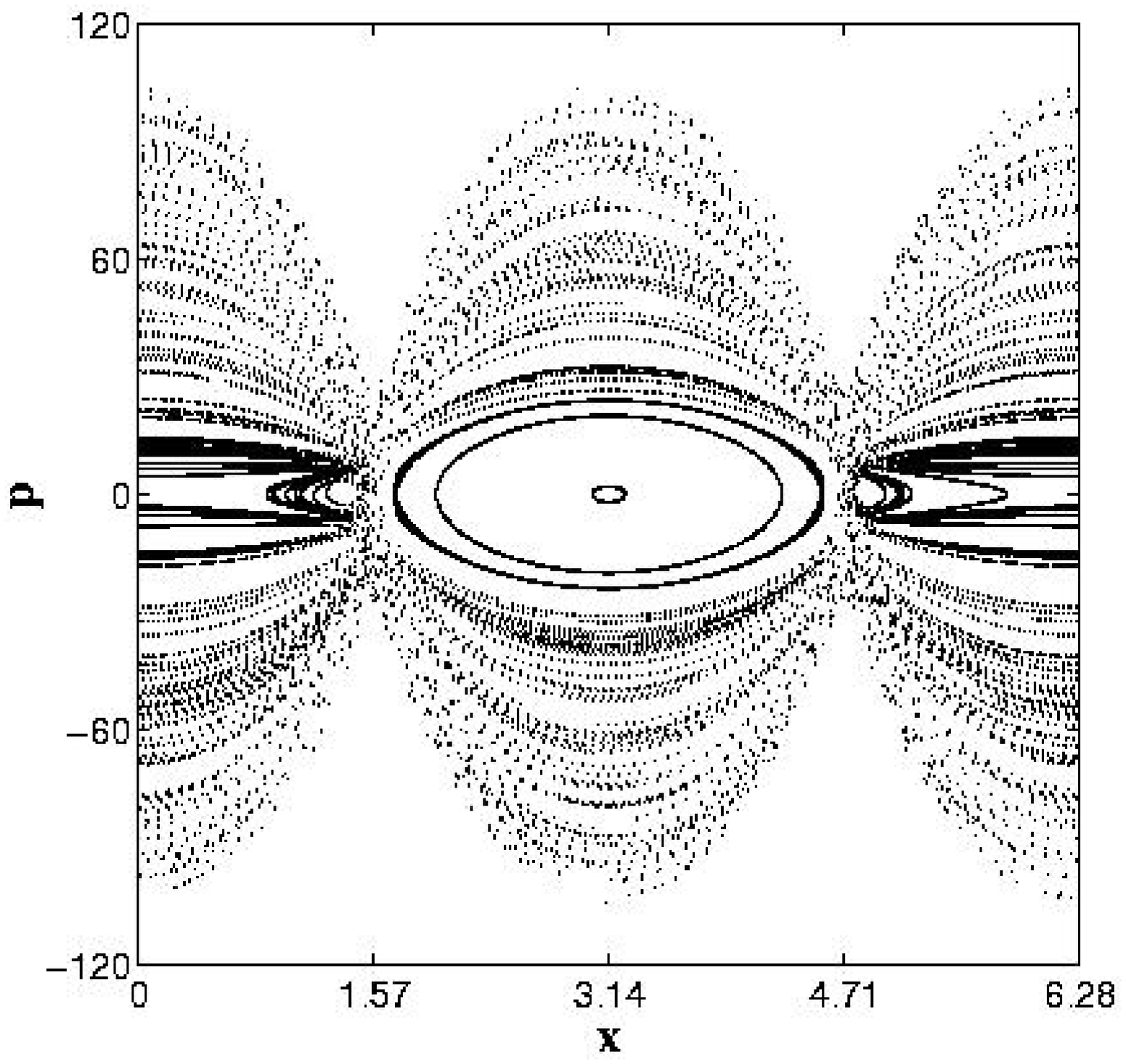}
\caption{\label{fig6}The same as in FIG.~\ref{fig2} but with $p(0) = 2$, $\delta = 0.4$ and
the atom initially prepared nearly in its excited state with two slightly different
values of the population inversion. (a) $s_z (0) = 0.863$. (b) $s_z (0) =
0.8660254$ (for comparison, additional trajectories with $p(0) > 2$ are 
shown). $x, p$ are dimensionless.}
\end{figure}

The main way of controlling the properties of $G(x,\tau;\nu )$ is to change
the system's topology in phase space.
Speaking about the topology, we have in mind the phase pattern (see also
the end of the \setcounter{abcd}{2}Sec.~\Roman{abcd}) that includes the
singular points, curves, and partitioning of the domains of chaos and
islands.
To illustrate how the system is sensitive to small variations of the initial
conditions that change the full energy, we show in FIGs.~\ref{fig6}a and b the Poincar\'{e} sections
with $s_z (0) = 0.863$ and $s_z (0) = 0.8660254$, respectively, at
$\delta = 0.4$ and under the other equal conditions. Very small difference
in the values of the initial tipping angle between the direction of the Bloch
vector and the axis $z$ gives rise to cardinally different motion with 
$p(0) = 2$, chaotic
oscillations in the wide range of the atomic momenta with $s_z (0) = 0.863$
and small regular translational oscillations nearly the bottom of a potential
well with $s_z (0) = 0.8660254$. Transition from order to chaos takes place
with the latter value of the initial atomic population inversion only with
much more large values of the initial momentum,
$p(0)\gtrsim 40$.

\begin{figure}[!htb]
\includegraphics[width=0.23\textwidth,clip]{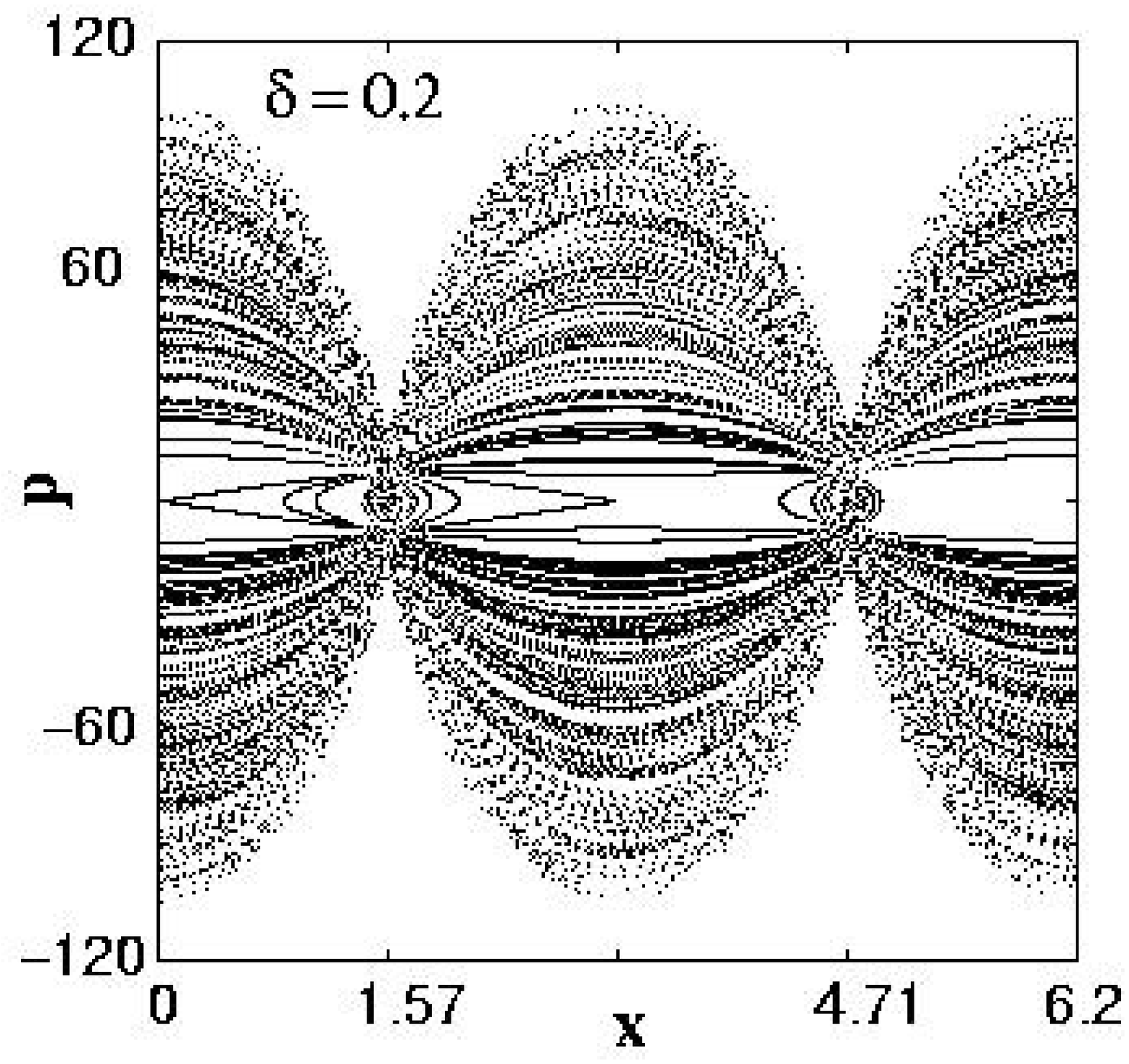}\hfill
\includegraphics[width=0.23\textwidth,clip]{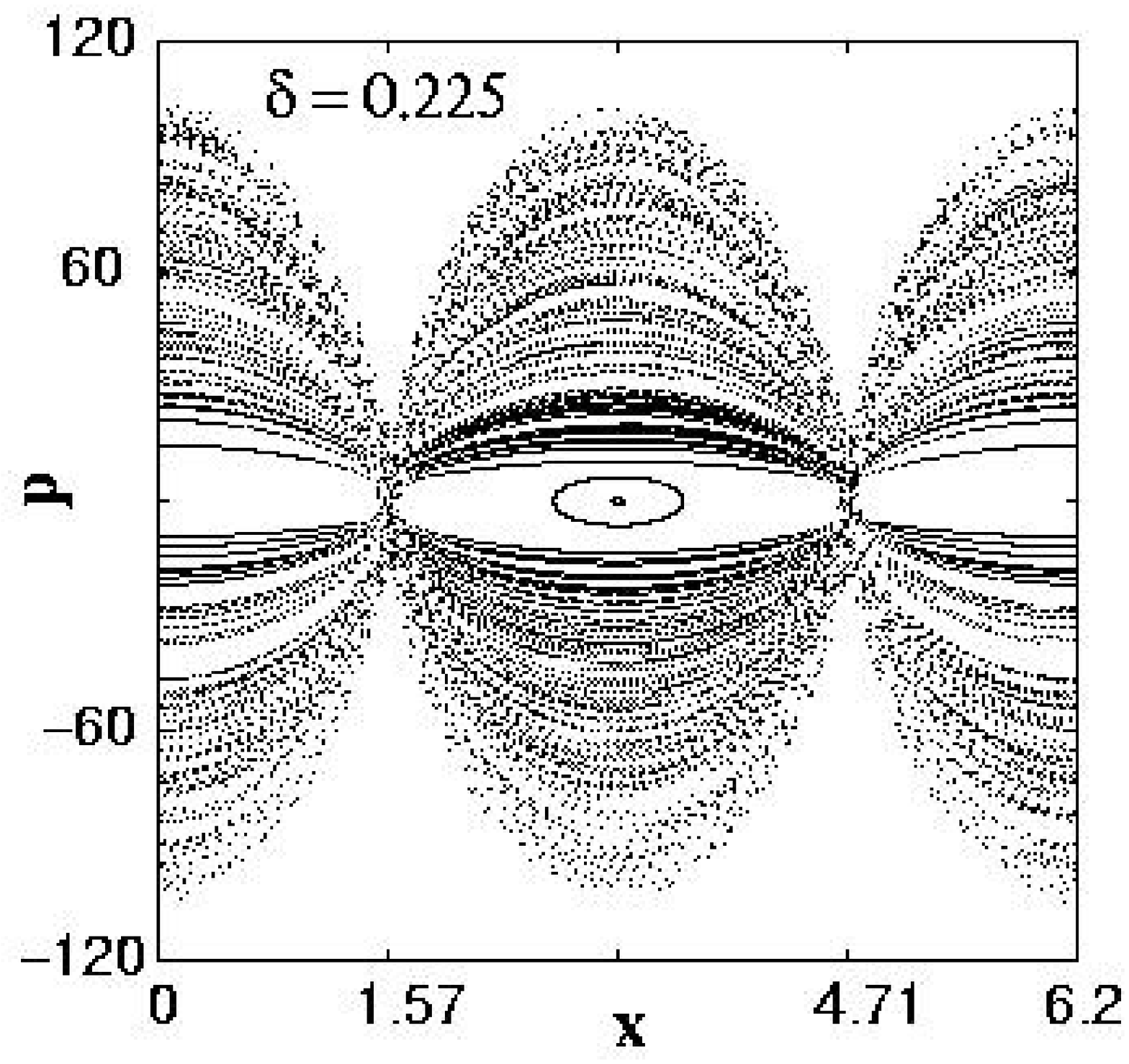}
\caption{\label{fig7}Bifurcation of the saddle point into the elliptic point on the plane
$x - p$ under changing the detuning $\delta$. $x, p$ are dimensionless.}
\end{figure}

The original system (\ref{2.3}) has three degrees of
freedom and it is not studied yet. Nevertheless, we were able  
to demonstrate by simulation a bifurcation of
a hyperbolic point into the elliptic one, although we are not able to
provide an analytical description at the moment since the   FIG.~\ref{fig7}
is just a projection of a trajectory in the 4-dimensional hyperspace onto
the plane $(p,x)$. By a change of $\delta$ near
$\delta^* \sim 0.222$, the saddle on the $x - p$ plane with $a_x =s_x = 0$
transforms into the elliptic point. The trapping potential well of the
finite size on $x - p$ plane occurs as a result of the bifurcation: this
bifurcation will be studied in detail in another paper.

\section{Conclusion}

A system with one or more cold atoms strongly coupled to a single mode
of the cavity field is ideal for testing fundamentals of quantum mechanics
and its corresponding to classical mechanics. Based on our understanding
of the nonlinear dynamics of the atom-photon interaction in a standing-wave
high-finesse cavity, new ways to manipulate and control atomic motion
can be opened. We have shown that the motion is very sensitive to the
atom-cavity detuning $\delta$. Varying  $\delta$, one can design topology
of the underlying phase space creating zones of trapping, quasi-trapping or
acceleration, quasi-regular  and stochastic webs, etc. It may provide new schemes
for cooling, trapping, and accelerating atoms.

To give an idea about the values of the magnitudes we have used in
numerical simulations, we need to estimate the range of values of the
normalized recoil frequency $\alpha = \hbar k_f^2 /m\Omega_0$ with
real atoms and cavities. We will use the parameters of the real experiments with single
atoms in the strong-coupling regime \cite{Y99,MF99},
for which the maximal atom-field coupling strength $\Omega_0$ exceeds the
decay rates of the cavity field and of the atomic dipole. Atoms were
collected in a magneto-optical trap and cooled down to microKelvin
temperatures, before entering a microscopic high-finesse Fabry-Perot cavity
with $Q \simeq 10^6$, $\Omega_0 \simeq 2\pi (10^7 - 10^8)$ Hz and
$k_f \simeq 2\pi
\cdot 10^6 m^{-1}$. With these values of the parameters, one can estimate
$\alpha$ to be in the range $10^{-5} - 10^{-2}$ depending on the atomic
mass and $\Omega_0$.

\section{Acknowledgment}

We thank L.E. Kon'kov for his help in preparing FIG.~\ref{fig1}. This work was
supported by the Russian Foundation for Basic Research under Grant No.
02-02-17796;
the Office of Naval Research, Grants No. N00014-96-1-0055 and No. N00014-97-1-0426,
and U.S. Department of Energy Grant No. DE-FG02-92ER54184. The
simulations were supported by the National Science Foundation cooperative
agreement No. AC1-9619020 through computing resources provided by the
National Partnership for Advanced Computational Infrastructure at the
San Diego Supercomputer Center.
S.P. thanks the Courant Institute of Mathematical Sciences, New York
University, for the hospitality during his stay.

\end{document}